\RequirePackage{amsthm}

\documentclass[pdflatex,sn-nature]{sn-jnl}

\usepackage{graphicx}%
\usepackage{amsmath,amssymb,amsfonts}%
\usepackage{amsthm}%
\usepackage{mathrsfs}%
\usepackage[title]{appendix}%
\usepackage{xcolor}%
\usepackage{textcomp}%
\usepackage{manyfoot}%
\usepackage{booktabs}%
\usepackage{algorithm}%
\usepackage{algorithmicx}%
\usepackage{algpseudocode}%
\usepackage{listings}%
\usepackage[noabbrev,capitalise,nameinlink]{cleveref}
\usepackage{gensymb}
\usepackage{lmodern}
\usepackage{anyfontsize}
\usepackage[version=4]{mhchem}
\usepackage{multicol}
\usepackage[left]{lineno}
\usepackage{multirow}
\usepackage{hhline}
\usepackage{setspace}

\theoremstyle{thmstyleone}%
\theoremstyle{thmstyletwo}%

\theoremstyle{thmstylethree}%

\begin{document}

\title[Article Title]{
Nanoscale Ultrafast Lattice Modulation with Hard X-ray Free Electron Laser
}


\author[1,3]{\fnm{Haoyuan} \sur{Li}}\email{hyli16@stanford.edu}
\author[2]{\fnm{Nan} \sur{Wang}}\email{nanw0321@stanford.edu}
\author[5,12,13]{\fnm{Leon} \sur{Zhang}}\email{zleon@stanford.edu}
\author[3]{\fnm{Sanghoon} \sur{Song}}\email{sanghoon@slac.stanford.edu}
\author[3]{\fnm{Yanwen} \sur{Sun}}\email{yanwen@slac.stanford.edu}
\author[3]{\fnm{May-Ling} \sur{Ng}}\email{mlng@slac.stanford.edu}
\author[3]{\fnm{Takahiro} \sur{Sato}}\email{takahiro@slac.stanford.edu}
\author[4]{\fnm{Dillon} \sur{Hanlon}}\email{dfhanlon@asu.edu}
\author[4]{\fnm{Sajal} \sur{Dahal}}\email{sdahal15@mainex1.asu.edu}
\author[3]{\fnm{Mario D.} \sur{Balcazar}}\email{mdbm@slac.stanford.edu}
\author[3]{\fnm{Vincent} \sur{Esposito}}\email{vincent.esposito@stanford.edu}
\author[5]{\fnm{Selene} \sur{She}}\email{bshe@stanford.edu}
\author[6,12]{\fnm{Chance} \sur{Caleb Ornelas-Skarin}}\email{ccornela@stanford.edu}
\author[7]{\fnm{Joan} \sur{Vila-Comamala}}\email{joan.vila-comamala@psi.ch}
\author[7]{\fnm{Christian} \sur{David}}\email{christian.david@psi.ch}
\author[8]{\fnm{Nadia} \sur{Berndt}}\email{nadiaber@mit.edu}
\author[8]{\fnm{Peter Richard} \sur{Miedaner}}\email{miedaner@mit.edu}
\author[8]{\fnm{Zhuquan} \sur{Zhang}}\email{zhuquan@mit.edu}
\author[1,9,11]{\fnm{Matthias} \sur{Ihme}}\email{mihme@stanford.edu}
\author[12, 13]{\fnm{Mariano} \sur{Trigo}}\email{mtrigo@slac.stanford.edu}
\author[8]{\fnm{Keith A.} \sur{Nelson}}\email{kanelson@mit.edu}
\author[3,9,12]{\fnm{Jerome B.} \sur{Hastings}}\email{jbh@slac.stanford.edu}
\author[8]{\fnm{Alexei A.} \sur{Maznev}}\email{alexei.maznev@gmail.com}
\author[10]{\fnm{Laura} \sur{Foglia}}\email{laura.foglia@elettra.eu}
\author[4]{\fnm{Samuel} \sur{Teitelbaum}}\email{SamuelT@asu.edu}
%\orgaddress{\street{367 Panama Street}, \city{Stanford}, \postcode{94305}, \state{CA}, \country{US}}
\author[5,9,12]{\fnm{David A.}\sur{Reis}}\email{dreis@stanford.edu}
\author*[3]{\fnm{Diling} \sur{Zhu}}\email{dlzhu@slac.stanford.edu}

\affil[1]{\orgdiv{Mechanical Engineering Department}, \orgname{Stanford University}, \orgaddress{\street{440 Escondido Mall}, \city{Stanford}, \postcode{94305}, \state{CA}, \country{US}}}

\affil[2]{\orgdiv{Physics Department}, \orgname{Stanford University}, \orgaddress{\street{382 Via Pueblo Mall}, \city{Stanford}, \postcode{94305}, \state{CA}, \country{US}}}

\affil[3]{\orgdiv{Linac Coherent Light Source}, \orgname{SLAC National Accelerator Laboratory}, \orgaddress{\street{2575 Sand Hill Rd}, \city{Menlo Park}, \postcode{94025}, \state{CA}, \country{US}}}

\affil[4]{\orgdiv{Physics Department}, \orgname{Arizona State University}, \orgaddress{\street{550 E Tyler Drive}, \city{Tempe}, \postcode{85287}, \state{AZ}, \country{US}}}

\affil[5]{\orgdiv{Applied Physics Department}, \orgname{Stanford University}, \orgaddress{\street{348 Via Pueblo}, \city{Stanford}, \postcode{94305}, \state{CA}, \country{US}}}

\affil[6]{\orgdiv{Electric Engineering Department}, \orgname{Stanford University}, \orgaddress{\street{350 Jane Stanford Way}, \city{Stanford}, \postcode{94305}, \state{CA}, \country{US}}}

\affil[7]{\orgname{Paul Scherrer Institute}, \orgaddress{\street{Forschungsstrasse 111}, \city{Villigen}, \postcode{5232 }, \country{Switzerland}}}

\affil[8]{\orgdiv{Department of Chemistry}, \orgname{Massachusetts Institute of Technology}, \orgaddress{\street{77 Massachusetts Ave}, \city{Cambridge}, \postcode{02139}, \state{MA}, \country{US}}}

\affil[9]{\orgdiv{Department of Photon Science}, \orgname{SLAC National Accelerator Laboratory}, \orgaddress{\street{2575 Sand Hill Rd}, \city{Menlo Park}, \postcode{94025}, \state{CA}, \country{US}}}

\affil[10]{\orgname{Elettra - Sincrotrone Trieste S.C.p.A.}, \orgaddress{\city{Basovizza}, \postcode{34149}, \country{Italy}}}

\affil[11]{\orgname{Department of Energy Science \& Engineering}, \orgname{Stanford University}, \orgaddress{\street{367 Panama Street}, \city{Stanford}, \postcode{94305}, \state{CA}, \country{US}}}

\affil[12]{\orgname{Stanford PULSE Institute}\orgaddress{\street{2575 Sand Hill Rd}, \city{Menlo Park}, \postcode{94025}, \state{CA}, \country{US}}}

\affil[13]{\orgname{Stanford Institute for Materials and Energy Sciences} \orgaddress{\street{2575 Sand Hill Rd}, \city{Menlo Park}, \postcode{94025}, \state{CA}, \country{US}}}

%\affil[13]{\orgdiv{Applied Physics Department}, \orgname{Stanford University}, \orgaddress{\street{348 Via Pueblo Mall}, \city{Stanford}, \postcode{94305}, \state{CA}, \country{US}}}


\abstract{
a    
}


\maketitle

\newpage

\textbf{
Understanding and controlling microscopic dynamics across spatial and temporal scales has driven major progress in science and technology over the past several decades.
While ultrafast laser-based techniques have enabled probing nanoscale dynamics at their intrinsic temporal scales down to femto- and attoseconds, the long wavelengths of optical lasers have prevented the interrogation and manipulation of such dynamics with nanoscale spatial specificity. 
With advances in hard X-ray free electron lasers (FELs) \cite{emma2010first, pile2011first, weise2017commissioning, milne2017swissfel}, 
significant progress has been made developing X-ray transient grating (XTG) spectroscopy\cite{rouxel2021hard, peters2023hard}, aiming at the coherent control of elementary excitations with nanoscale X-ray standing waves.
So far, XTGs have been probed only at optical wavelengths \cite{rouxel2021hard, peters2023hard}, thus intrinsically limiting the achievable periodicities to several hundreds of nm.
By achieving sub-femtosecond synchronization of two hard X-ray pulses at a controlled crossing angle, we demonstrate the generation of an XTG with spatial periods of $\sim10$~nm. 
The XTG excitation drives a thermal grating that drives coherent monochromatic longitudinal acoustic phonons in the cubic perovskite, \ce{SrTiO3} (STO).
With a third X-ray pulse with the same photon energy, 
time-and-momentum resolved measurement of the XTG-induced scattering intensity modulation provides evidence of ballistic thermal transport at nanometer scale in STO.
These results highlight the great potential of XTG for studying high-wave-vector excitations and nanoscale transport in condensed matter, and establish XTG as a powerful platform for the coherent control and study of nanoscale dynamics.
}

From the dynamic organization of cell membranes \cite{simons2011membrane} to the polarization and magnetization switching in next-generation memory devices \cite{meena2014overview}, the nanoscale dynamics of electrons and atoms underpins a wide range of essential functionalities in both natural and technological systems.
Significant advancements in nano-fabrication \cite{cui2008nanofabrication}, atomic-resolution microscopy \cite{krivanek2010atom}, and ultrafast spectroscopy \cite{kukura2007femtosecond} have greatly improved our ability to construct and probe nanoscale systems. 
To achieve coherent control of the dynamics on their intrinsic temporal and spatial scales, frequency-selective excitation techniques have been developed using resonant laser couplings with spin, charge, and lattice degrees of freedom \cite{teitelbaum2018frequency, li2019terahertz, kogar2020light, salikhov2023coupling}.
However, due to the long wavelengths of conventional lasers ranging from a few hundred nanometers to a few micrometers, these optical techniques inherently lack nanoscale spatial specificity.
It remains a major challenge to selectively induce dynamics with predefined nanoscale spatial profiles.

X-ray pulses provide a distinct approach to investigate microscopic dynamics with nanoscale spatial specificity, enabled by their intrinsic short wavelengths.
For decades, atomic-scale X-ray standing wave techniques \cite{bedzyk2002x} have been used to study impurities in bulk crystals and adsorbate structures on surfaces and interfaces through fluorescence spectroscopy.
The rapid advancement of ultrafast high brightness X-ray free electron lasers \cite{emma2010first, pile2011first, milne2017swissfel, weise2017commissioning} (FEL) offers the opportunity to utilize X-ray standing waves not only to probe but also to derive nanoscale dynamics.
Through near-instantaneous energy deposition with the same spatial periodicity as the X-ray standing wave, one could selectively address targeted degrees of freedom at nanometer or atomic length scales, thereby obtaining wavevector-specific control of matter.

\begin{figure}[hht!]
\centering
\includegraphics[width=0.8\textwidth]{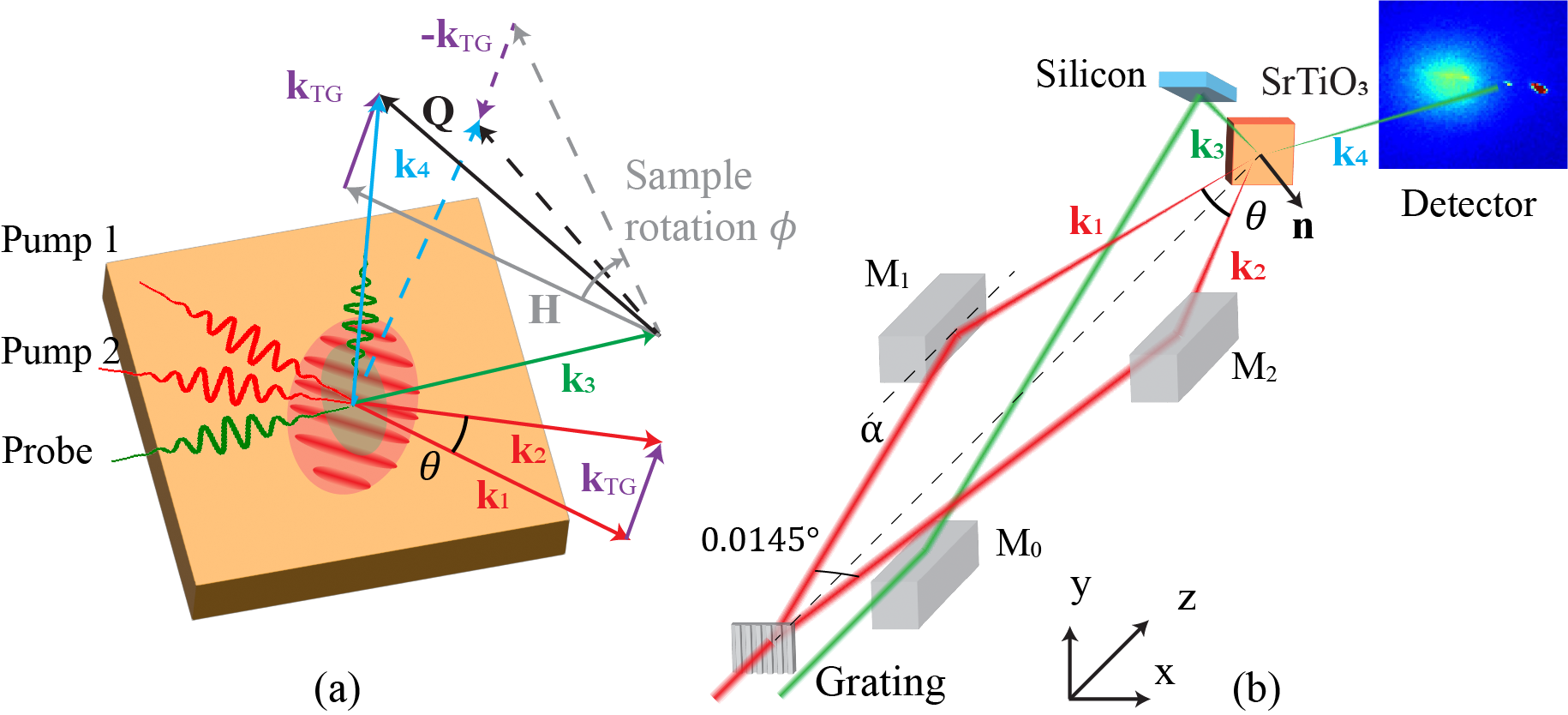}
\caption{\textbf{Schematics of the experiment}:
(a) Schematic illustration of XTG.
The two pump pulses (red), overlap on the sample in space and time with a crossing angle $\theta$. 
The probe pulse (green) overlaps with the pump pulses in space with a delay time $\tau$.
The wave vectors $\mathbf{k}_1$, $\mathbf{k}_2$ and $\mathbf{k}_3$ are associated with pump and probe pulses, respectively, and $\mathbf{k}_4$ is the scattered photon wavevector.
The XTG wavevector is $\textbf{k}_\text{TG} \equiv \textbf{k}_1 - \textbf{k}_2$,
and $\mathbf{Q}\equiv \textbf{k}_4 - \textbf{k}_3$ indicates the wave-vector change of the scatterred probe photons. 
$\textbf{H}$ represents the reciprocal lattice vector of STO (220) Bragg reflection.
(b) Experimental setup.
The pump pulse is diffracted by the diamond transmission grating and the $\pm1$ diffraction orders are recombined on the sample through total reflections with Rh-coated mirrors, $M_1$ and $M_2$.
The probe pulse is overlapped with the pump pulses with the total reflection with the Rh-coated mirror, $M_0$, and a silicon Bragg mirror with (111) symmetric Bragg reflection. 
The sample is a STO crystal, which can rotate around its normal direction, $\textbf{n}$. 
The area detector records X-ray scattering from the sample.
The split-delay optics (SDO) that generates the pump and probe pulses is not shown.
}\label{fig1_setup}
\end{figure}

In the technique referred to as X-ray transient grating (XTG)\cite{rouxel2021hard}, the periodic excitation pattern could diffract the probe pulse like a physical grating, as shown in \cref{fig1_setup}(a).
Efforts to realize nanoscale XTG have persisted over the past decade, driven by its potential in nonlinear optics \cite{chergui2023progress} and ultrafast nanoscale control of microscopic processes, such as spin and charge patterns in quantum materials \cite{ambrosetti2016wavelike, miedaner2024excitation}, or nanoscale heterogeneity in ionic liquids \cite{hayes2015structure}.
While an analogous approach with extreme ultraviolet (EUV) lasers  \cite{bencivenga2019nanoscale} has been demonstrated around the same time to study nonlinear optics and nanoscale dynamics in thin films \cite{foglia2023extreme, miedaner2024excitation, bencivenga2015four, foglia2018first}, so far XTG has been only demonstrated with periodicity $\Lambda_\text{TG}$ on the sub-1~$\mu$m scale with an optical probe pulse \cite{rouxel2021hard, peters2023hard}.
Furthermore, a key challenge lies in the X-ray-matter interaction mechanism.
%deposit energy primarily through single X-ray photon absorption, 
Hard X-ray photons produce keV photoelectrons with an inelastic mean free path in solids on the order of 10~nm \cite{de2019electron}.
This may lead to spatially extended energy deposition footprints and thus smear out the periodic structure of XTG excitations on comparable length scales.
While recent studies provide evidence for X-ray-induced nanoscale lattice modulation \cite{huang2024Nanometer}, it has been an open question whether XTG could effectively create periodic perturbations on 10~nm and smaller scales.

We report an ultrafast XTG with 10~nm spatial periodicity, using 9.8~keV X-ray laser pulses for both excitation and detection.
The XTG induces periodic lattice modulations in a bulk SrTiO$_3$ (STO) crystal, which manifest as a sharp satellite transient X-ray diffraction peak atop an otherwise smoothly varying diffuse scattering pattern generated by a time-delayed X-ray probe pulse.
The peak intensity oscillates in time at both the fundamental and second harmonic frequency corresponding to longitudinal acoustic (LA) phonons with wavelength matching the TG period.
Analysis of the time-dependent XTG diffraction intensity reveals that the lattice modulation consists of a thermal grating and a standing strain wave. 
Our results suggest ballistic thermal transport on a 5.9~nm length scale in bulk STO.
The demonstrated capability to create spatially periodic excitations with nanometer spatial periods establishes XTG as a powerful tool for coherently controlling nanoscale dynamics.

The experiment is conducted at the XPP instrument at LCLS \cite{chollet2015x}.
The setup is illustrated in \cref{fig1_setup}(b),
where both the pump (red) and probe (green) pulses have a wavelength of $\lambda_0=0.127~$nm and a pulse duration of 30~fs.
They are generated from a single incident X-ray pulse using  split-delay optics (SDO) \cite{li2021generation}, which allows an adjustable delay time between -1 and 14~ps (see \ref{fig:whole setup} and Methods for details).
The X-ray pulse has a coherence time of 10~fs corresponding to a bandwidth of 0.4~eV \cite{li2021generation}.
The initial pump pulse is split with a transmission grating.
Its $\pm1$ diffraction orders are brought back onto the sample using a pair of Rh-coated grazing-incidence mirrors.
To generate high-contrast nanoscale interference fringes across the full pulse, the two crossing pump pulses must have a high mutual spatial coherence and an arrival time difference shorter than the coherence time.
This is achieved through a symmetric optical layout as shown in \cref{fig1_setup}(b) and precise alignment of $M_1$ and $M_2$, with the same grazing incidence angle, $\alpha$.
An accuracy of $0.003\degree$ of $\alpha$ leads to sub-femtosecond-precision synchronization.
The intensity distribution of the resulting X-ray interference pattern has a spatial period, $\Lambda_\text{TG} = \lambda_0 \left(2 \sin\left(\theta /2\right)\right)^{-1}$, where $\theta$ is the crossing angle between the two pump pulses.
We used two configurations with $\theta$ set to $0.615\degree$ and $0.785\degree$, corresponding to $\Lambda_\text{TG}=11.8$ and $9.2$~nm respectively.
We define the XTG wavevector $\textbf{k}_\text{TG} \equiv \textbf{k}_1 - \textbf{k}_2$, noting that the standing wave comprises spatial periodicities at $\pm\textbf{k}_\text{TG}$.
With respect to the STO unit cell axes, $\textbf{k}_\text{TG}$ is oriented along $\left[-0.012, 0.980, 0.199\right]$ direction.

\begin{figure}[t!]
\includegraphics[width=0.98\textwidth]{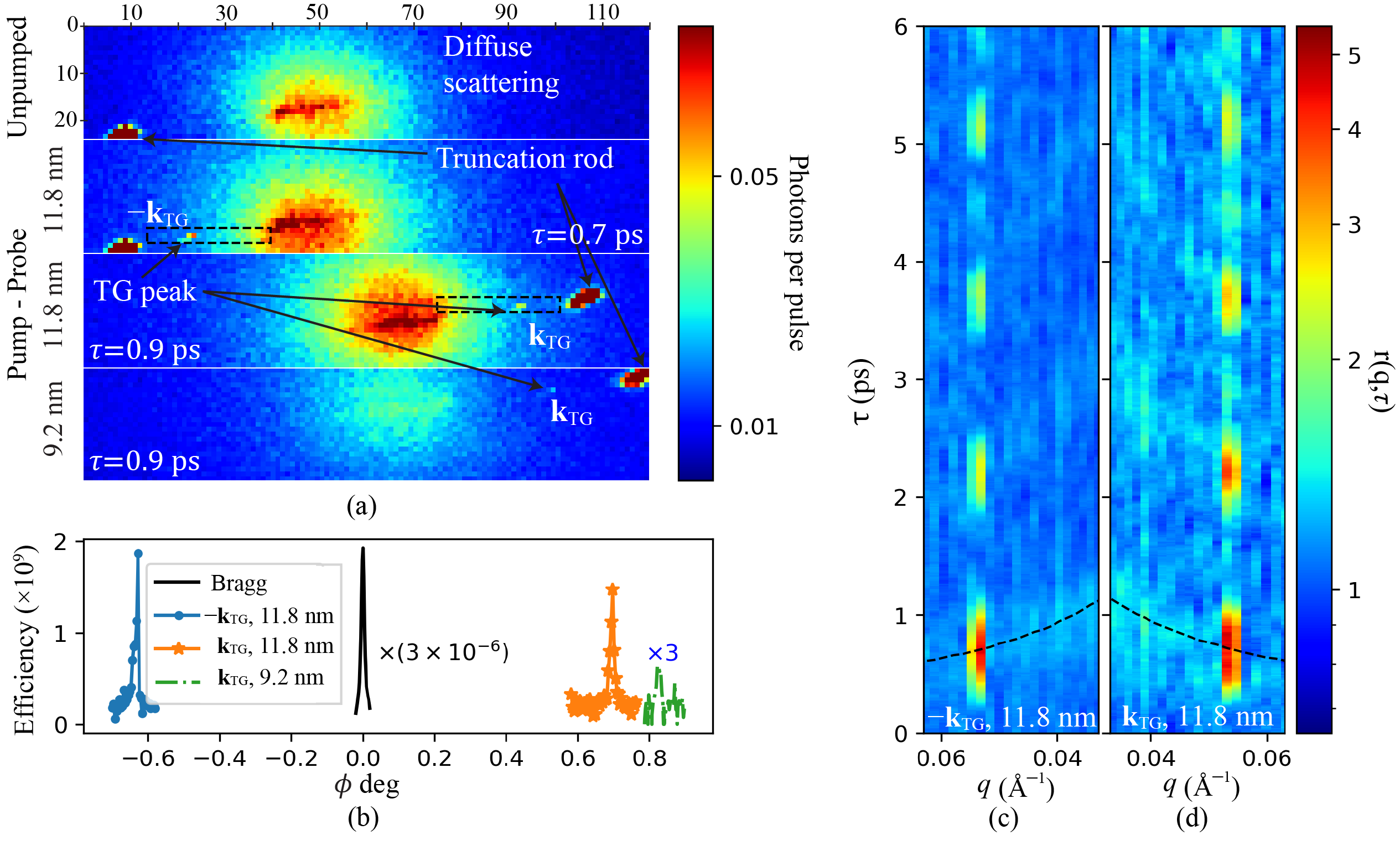}
\caption{\textbf{XTG scattering intensity}:
(a) Detector area where diffuse scattering and TG peaks are observed.
The color scale is the number of photons per pixel per pulse. 
The top panel is the diffuse scattering from the unpumped sample after rotating the sample by $\phi=-0.629\degree$ away from the Bragg peak.
The axes tickes are pixel indexes. 
The middle two panels are the scattering intensity with XTG pumps for $\Lambda_\text{TG}=11.8~nm$, respectively with $\phi=-0.629\degree$ and $\phi=0.700\degree$.
The second panel corresponding to $\textbf{q}=\textbf{k}_\text{TG}$, while thrid panel corresponding to $\textbf{q}=\textbf{k}_\text{TG}$.
The fourth panel corresponds to $\Lambda_\text{TG}=9.2~nm$ for $\textbf{q}=\textbf{k}_\text{TG}$ and $\phi=0.823\degree$.
The pump probe delay time $\tau$ is indicated in each panel.
(b) The angular dependence of the diffraction efficiency for the STO Bragg peak and TG peaks.  
(c) and (d) the intensity modulation, $r(\textbf{q},\tau) \equiv I(\textbf{q} + \textbf{H},\tau) / I(\textbf{q} + \textbf{H},0)$, for regions within dashed black boxes in (b), versus the pump probe delay time $\tau$.
The black dashed lines indicate the first maxima in $\tau$ for each $q$.
}\label{det_img_and_time_trace}
\end{figure}

The probe beam is delivered to the sample by reflection from the mirror $M_0$ followed by a Si(111) Bragg reflection, leading to a crossing angle of $23 \degree$ between the pump and probe pulses within the vertical plane.
The pump and probe pulse beam sizes at the sample are all $5\times10~\mu\text{m}^2$.
The average pulse energy on the sample is 1.6~$\mu$J for the probe pulse and 0.8~$\mu$J for each individual pulse in the pump pulse pair.
The sample is an STO single crystal with surface normal perpendicular to the (010) planes, and can be rotated by angle $\phi$ around its surface normal.
Both pump pulses impinge on the sample with an incident angle of approximately $11.5\degree$ with respect to the surface.
A 3D schematic of X-ray and sample geometry is shown in \ref{fig:whole setup} (e).
Scattered photons from both pump and probe pulses are recorded by an area detector.
Our geometry is close to the Bragg condition for the (220) reflection of the probe, in which case the contribution of the scattered pump pulses is negligible. 
XTG diffraction signal from the probe pulse is located when rotating STO azimuth away from the (220) Bragg reflection.

\cref{det_img_and_time_trace} (a) shows snapshots of the measured scattering intensity near the (220) Bragg reflection of STO (220) for 3 different sample angles $\phi$ away from the Bragg peak, where $\phi=0$ corresponds to the Bragg reflection. 
Scattered probe photons measured by each pixel have a wavevector change $\textbf{Q}\equiv \mathbf{k}_\text{4}-\mathbf{k}_\text{3}$.
At each sample angle $\phi$, the rotated STO (220) reciprocal lattice vector $\textbf{H}$ maps $\textbf{Q}$ to an equivalent reduced wave-vector in the first Brillouin zone (BZ), $\textbf{q}\equiv \mathbf{Q} - \mathbf{H}$.
The diffuse scattering shown in the top panel of \cref{det_img_and_time_trace}(a) arises from static defects and thermal fluctuations of atomic positions in STO.
Upon XTG excitation, a sharp diffraction spot emerges atop the diffuse background, as indicated by the black arrows in the lower three panels of \cref{det_img_and_time_trace}(a),  at $\phi=-0.629\degree$, $0.700\degree$, and $0.823\degree$.
The XTG peak is tightly localized around the pixel with $\textbf{q}\approx \pm \textbf{k}_\text{TG}$ for $\Lambda_\text{TG}=11.8$ and $9.2$~nm, unambiguously demonstrating that the X-ray standing wave has created lattice modulations with the corresponding spatial periodicity.

\begin{figure}[bht!]
\centering
\includegraphics[width=0.5\textwidth]{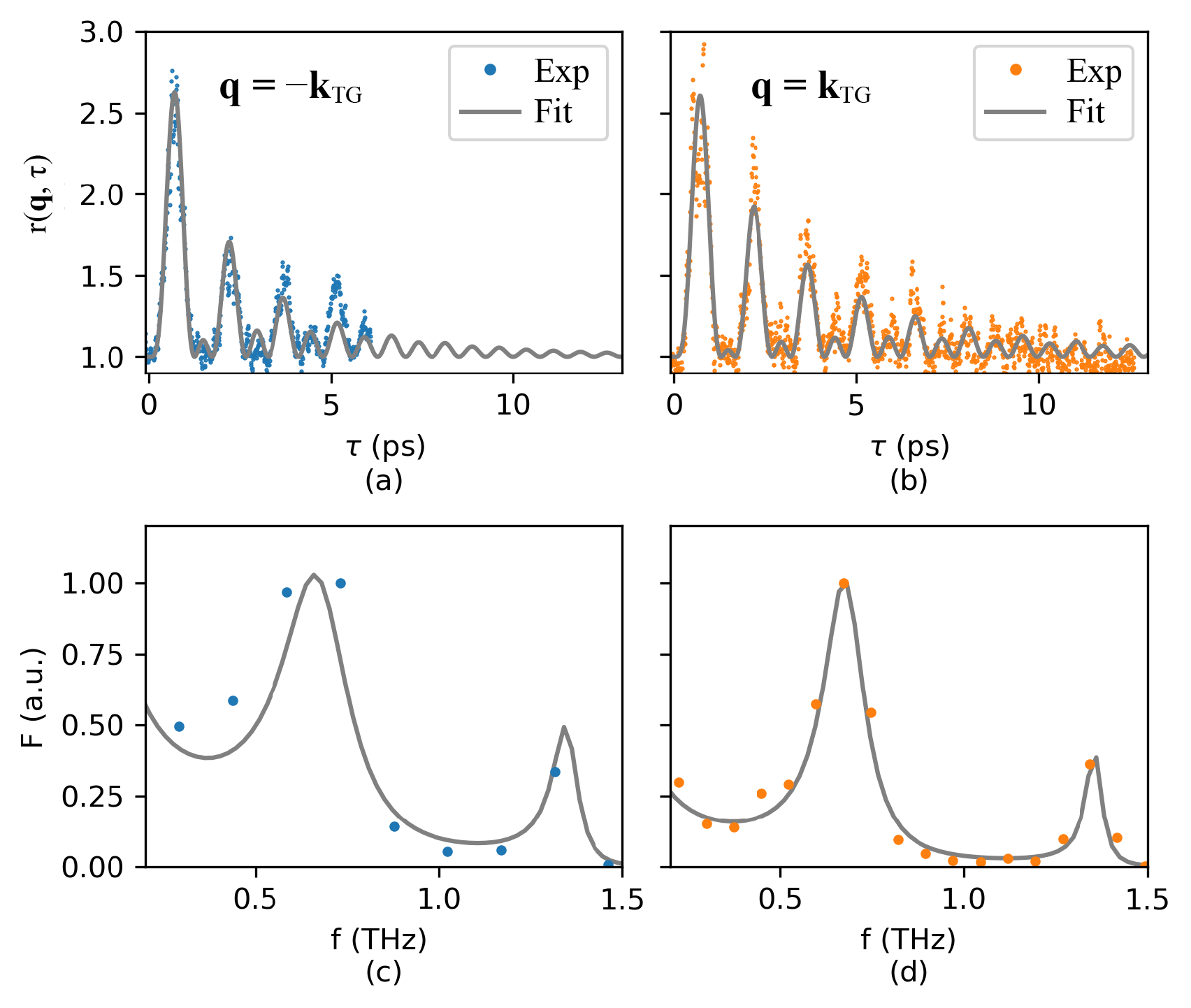}
\caption{\textbf{Time dependence and spectral composition of XTG signal}:
(a) and (b) are time traces of $r(\textbf{q},\tau)$ at angular peaks of XTG with $\Lambda_\text{TG}=11.8$~nm according to \cref{det_img_and_time_trace} (b).
The solid line is the fitting result with \cref{eq:final r}.
(c) and (d) are Fourier transforms of $r(\textbf{q},\tau)$ of (a) and (b) respectively.
The solid line is the Fourier transforms of the corresponding fitted curves in (a) and (b).
We have normalized all Fourier transformed curves to unity at $f=0.67~$THz.
}\label{time_trace_and_spec}
\end{figure}

The location of XTG peaks is consistent with the phase matching condition: $\textbf{k}_4 = \pm\left(\textbf{k}_1 - \textbf{k}_2\right) + \textbf{k}_3 + \textbf{H}$, as illustrated in \cref{fig1_setup}(a).
The presence of $\textbf{H}$ reflects the underlying crystalline symmetry and distinguishes this phase matching condition in the Bragg reflection geometry from that of optical transient grating measurements \cite{knoester1991transient}. 
The angular dependence of the diffraction efficiency of Bragg and XTG peaks is shown in \cref{det_img_and_time_trace}(b) for $\Lambda_\text{TG}=11.8~\text{and}~9.2~\text{nm}$.
All diffraction peaks exhibit a similar angular width of approximately $0.012\degree$, as illustrated in \ref{fig:area}.
The angular spread of the Bragg peak of $0.012\degree$ is broader than the ideal value of $0.003\degree$ with a perfect crystal according to the dynamical diffraction theory \cite{shvyd2004x}
(see \ref{fig:rocking_simulation} and Supplementary Note 3.1).
This reflects crystalline imperfections in the sample.
\cref{det_img_and_time_trace}(c) and (d) show time-dependent scattering intensity modulation $r(\textbf{q},\tau) \equiv I(\textbf{q} + \textbf{H},\tau) / I(\textbf{q} + \textbf{H},0)$ at various $\textbf{q}$ within the black boxes marked in \cref{det_img_and_time_trace}(a) for $\Lambda_\text{TG}=11.8~\text{nm}$.
\cref{time_trace_and_spec} shows the time trace and frequency spectra of $r(\textbf{q}, \tau)$ at $\textbf{q}=\pm \textbf{k}_\text{TG}$, reflecting the evolution of the XTG-induced periodic lattice modulation.
The signal exhibits oscillations at 0.67~THz and 1.34~THz, which coincide with the fundamental and second harmonic frequencies of LA phonons at $\pm \textbf{k}_\text{TG}$.
At longer delay times, the 1.34~THz component becomes more prominent.

Based on the pump pulse intensity and STO attenuation length at 9.8~keV, the average spacing between photo-absorption sites is approximately 13.5~nm.
This leads to a sparse and stochastic arrangement of photon-absorption sites across the X-ray illuminated volume.
Ultrafast X-ray diffuse scattering studies have shown that each X-ray photon-absorption event in STO gives rise to a nanoscale strain field \cite{huang2024Nanometer}, which is composed of a local thermal expansion and an outgoing 3D wave packet of LA phonons. 
In the wave-vector representation, the \textbf{q} component of the atomic displacement with a delay time $\tau$ associated with each single X-ray photon-absorption site is:
\begin{equation}
    \widetilde{\mathbf{u}}\left(\mathbf{q},\tau\right)=-\frac{u_0}{q}e^{-\frac{1}{2}\sigma^2q^2}\left(e^{-\frac{t}{\tau_0}}-e^{-\frac{t}{\tau_1}}\cos{\left(qvt\right)}\right)\hat{\mathbf{q}} , \label{eq:strain fields}
\end{equation}
where $\sigma$ measures the spatial extension of the initial lattice excitation, $u_0$ characterizes the displacement amplitude, $\tau_0$ represents the relaxation time of thermal grating, $\tau_1$ is the phonon lifetime, $v$ is the longitudinal sound velocity of STO along $\textbf{k}_\text{TG}$, and $\hat{\textbf{q}}\equiv \textbf{q} /q$.

We assume that there are $N$ photon-absorption sites within each XTG coherent scattering volume $V\equiv l^3$, where $l$ is the spatial extension of the coherent X-ray standing wave.
Denoting the center of each site with $\mathbf{r}_s$, $s=1,2,\ldots,N$,
the resulting intensity modulation is (see Supplementary Note 2 for derivations):
\begin{equation}
    r\left(\textbf{q},\tau \right) - 1 \propto \left\langle\left|\sum_{s=1}^{N}e^{i\mathbf{q}\cdot\mathbf{r}_s}\right|^2\right\rangle\left|\mathbf{Q}\cdot\tilde{\mathbf{u}}(\textbf{q},\tau)\right|^2 , \label{eq:diffuse scattering} 
\end{equation}
where $\langle \cdot \rangle$ indicates the ensemble average over independent pump-probe pulses, and over coherent scattering volumes $V$ within the total X-ray illuminated region. 
The distribution of photon-absorption sites follows the X-ray standing wave intensity profile $P(\textbf{r}) = V^{-1}\left(1+\cos{\left(\mathbf{r}\cdot\mathbf{k}_\text{TG}\right)}\right)$, where $\beta$ is the averaged contrast of the X-ray standing wave.
This probability distribution leads to
\begin{equation}
    \left\langle\left|\sum_{s=1}^{N} e^{i\mathbf{q}\cdot\mathbf{r}_s}\right|^2\right\rangle=\left\{\begin{matrix}N&for\ \ \mathbf{q}\neq\pm\mathbf{k}_{TG} \\ \frac{N\left(N+3\right)}{4}&for\ \ \mathbf{q}=\pm\mathbf{k}_{TG}\\
    \end{matrix}\right. .
\end{equation}
Therefore, for $\textbf{q}\neq \pm \textbf{k}_\text{TG}$, the intensity modulation results from an incoherent summation of individual photon-absorption sites.
In contrast, at $\textbf{q}= \pm \textbf{k}_\text{TG}$, the coherent superposition of scattering amplitude from $N$ sites leads to an enhancement factor of $(N +3)/4$.

This coherent enhancement of the scattering intensity at $\textbf{q} = \pm\textbf{k}_\text{TG}$ is clearly observed in \cref{det_img_and_time_trace}(c) and (d). 
The black dashed lines indicate the location of the maxima of $r(\textbf{q},\tau)$ in $\tau$ for each $\textbf{q}$.
Along these lines, $r(\textbf{q},\tau) - 1$ at 
$\textbf{q}=\pm\textbf{k}_\text{TG}$ exceeds the extrapolated baselines from neighboring $\textbf{q}\neq \pm \textbf{k}_\text{TG}$ by a factor of 11 and 17 respectively. (see \ref{fig:enhancement factor} and Supplementary Note 3.2 for details).
For a grating with a linear extension $l$ and a photon wavelength $\lambda_0$, the FWHM angular spread of its diffraction is $\delta\phi\approx 0.443 \times \lambda_0 \left(l \cos{\theta_\text{B}}  \cos{\eta}\right)^{-1}$.
Here, $\theta_\text{B}=27.270\degree$ is the Bragg angle and $\eta = 46.107\degree$ is the angle between $\textbf{n}$ and $\mathbf{\sigma}\equiv\frac{\mathbf{k}_3\times\mathbf{H}}{\left|\mathbf{k}_3\times\mathbf{H}\right|}$ (see details in the Supplementary Note 2.4).
For the measured angular spread of XTG peaks of $\delta \phi \approx 0.012\degree$, this leads to $l\approx 430$~nm.
Given an average distance of 13.5~nm between photon absorption sites, this corresponds to $N\approx 3.2\times10^4$ per coherent scattering volume $V=l^3$.
Taken into account the solid angle mismatch of the detector pixel, $\Omega_p=(0.026\degree)^2$, and the XTG diffraction peak, $\Omega_d\approx(0.012\degree)^2$, the theoretical estimation of the enhancement is $\Omega_d/\Omega_p \times (N+3)/4\approx 1800$.
The deviation from the theoretical estimation of the XTG coherent enhancement is primarily induced by the electron mean free path of 10~nm\cite{de2019electron}, which reduces the contrast of the periodic lattice modulation.

\begin{table}[bht!]
\begin{tabular}{|l|l|l|l|l|l|}
\hline
     & $\tau_0$ (ps) & $\tau_1$ (ps)   & $v$ (km/s)   \\ \hline
$-\textbf{k}_\text{TG}$  & $1.8\pm0.4$  & $7.7\pm1.4$ & $8.0\pm0.4$ \\ \hline
$+\textbf{k}_\text{TG}$  & $3.0\pm1.0$  & $11.9\pm3.6$ & $8.0\pm0.7$  \\ \hline
\end{tabular}
\caption{$\vert~$\textbf{Parameters of XTG-induced dynamics}: Lifetimes and sound velocities derived from measurements shown in \cref{time_trace_and_spec} with \cref{eq:final r}. 
The uncertainty reported here is the numerical output from of the fitting algorithm.
}\label{table:TG fitting main}
\end{table}

Fitting the measurements in \cref{time_trace_and_spec} with
%For $\textbf{q}=\pm \textbf{k}_\text{TG}$, 
\begin{equation}
    r\left(\pm \textbf{k}_\text{TG},\tau \right) - 1 \propto \beta^2 \frac{N\left(N-1\right)}{4}\left(e^{-\frac{\tau}{\tau_0}}-e^{-\frac{\tau}{\tau_1}}\cos{\left(k_\text{TG}v\tau\right)}\right)^2 \label{eq:final r}.
\end{equation}
we derive the longitudinal sound velocity, $v=8.0$ km/s, in good agreement with the macroscopic value of 7.8 km/s \cite{ishidate1988brillouin}.
The extracted $\tau_0$ and $\tau_1$ are summarized in \cref{table:TG fitting main}, and their difference leads to the time trace shape in \cref{time_trace_and_spec}:
initially, $r(\pm\textbf{k}_\text{TG}, \tau)$ is dominated by the cross term in $e^{-\frac{\tau}{\tau_0}-\frac{\tau}{\tau_1}}\cos{\left(\tau k_\text{TG}v\right)}$;
over time, thermal equilibration establishes faster than phonon relaxation, leaving behind only the second harmonic oscillations as seen in the measurement.
While the detector pixel solid angle $\Omega_p$ is larger than the XTG peak solid angle $\Omega_d$, the frequency span associated with $\Omega_d$ is 6~GHz, corresponding to a dephasing time of 58~ps.
Therefore, the measured $\tau_1$ reflects the actual LA phonons lifetime associated with the XTG-induced lattice modulation, and is also compatible with previous time-resolved X-ray diffuse scattering studies \cite{shayduk2022femtosecond}.

%the measured $\tau_1$ is influenced by both the intrinsic phonon lifetime and the dephasing of phonon modes with different frequencies.
%Considering the XTG peak angular span of $\delta\phi\approx 0.012\degree$, the longitudinal sound speed of $v=8.0\pm0.4$~km/s leads to a frequency span of 6~GHz, resulting in a dephasing time of $\tau_1^\prime \approx 13$~ps.
%Therefore, $\tau_1$ is primarily dominated by the dephasing of XTG-induced LA phonons.

%To reach thermal equilibrium within $\tau_0$, heat should move over $\Lambda_\text{TG}/2=5.9$~nm, at an effective velocity of 2.5 km/s.
%This high effective heat transport speed suggests that a significant fraction of the heat-carrying phonons propagate ballistically on this length scale \cite{maznev2011onset}.
%Indeed, according to the heat diffusion equation, the thermal grating decay time is given by $\tau_0 =(D k_\text{TG}^2)^{-1}$ \cite{eichler2013laser}.
%With the literature value of the thermal diffusivity of STO \cite{martelli2018thermal}, $D=0.04~\text{cm}^2/s$, this yields $\tau_0=0.9$~ps.
%This would imply that heat moves at an effective velocity of 6.7~km/s, faster than the acoustic mode mean group velocity of 5.24~km/s of STO at 300~K  \cite{fumega2020understanding}.
%Thus, these observations indicate that at the 6~nm length scale, the heat diffusion model breaks down, and heat transport enters the ballistic regime. 
%Our measurement opens possibilities to map the transition from diffusive to ballistic thermal transport on the sub-10~nm length scale in bulk solids.

Given the thermal relaxation time of $\tau_0$, we extract an effective thermal diffusivity \cite{eichler2013laser} $D_\text{eff}=\left(\tau_0k_\text{TG}^2\right)^{-1}=0.015\pm0.002$~$\text{cm}^2$/s, which is approximately 3 times smaller than the bulk value for STO, $D=0.04~\text{cm}^2/s$ \cite{martelli2018thermal}.
This reduced $D_\text{eff}$ suggests that heat transport in STO on the sub-10~nm scale deviates from conventional diffusion and may enter the ballistic regime, where effective diffusivity is reduced due to the limited phonon scattering on this length scale \cite{eichler2013laser}.
For comparison, consider a diffusive transport model with $D$.
To reach equilibrium over $\Lambda_\text{TG}/2=5.9$~nm within $\tau_0^\prime=(D k_\text{TG}^2)^{-1}=0.9$~ps, heat transfer should have an effective velocity of 6.7~km/s, exceeding the acoustic mode mean group velocity of 5.24~km/s of STO at 300~K \cite{fumega2020understanding}.
Since heat transfer velocity in STO should not supersede the group velocity of heat-carrying phonons, which are the acoustic phonons, this invalidates the diffusion model with $D$.
Thus, our results suggest ballistic heat transport in STO at the 6~nm length scale, and open possibilities to perform sub-10~nm transport studies in bulk solids.

Our results are the first demonstration of imposing periodic lattice modulations and exciting coherent collective motions in bulk crystals on the 10~nm length scale with XTG.
From the relaxation dynamics of the associated lattice modulation, we provide evidence of ballistic thermal transport in bulk STO on sub-10~nm length scales.
The ability of XTG to induce nanoscale periodic strain offers opportunities for studying the coupling between strain fields and nanoscale spin and charge waves \cite{ambrosetti2016wavelike, miedaner2024excitation}, nano-domains in multiferroics \cite{jia2015nanodomains}, and nonlinear lattice dynamics \cite{kozina2019terahertz}.
In addition, this nanoscale XTG utilizes all hard X-ray four-wave mixing and lays the groundwork for hard X-ray multidimensional spectroscopies \cite{chergui2023progress}.
With the advances in high-repetition hard X-ray FELs \cite{decking2020mhz, raubenheimer2018lcls} and multicolor \cite{hara2013two, lutman2013experimental} hard X-ray pulses, 
one may exert and observe spatially periodic and element-selective lattice modulations through a combination of core-level transitions.
Thus, the demonstrated XTG technique represents a critical step toward harnessing core-level transitions for coherent excitation and control of nanoscale dynamics  \cite{chergui2023progress}.

%\end{document}

\newpage 

%%%%%%%%%%%%%%%%%%%%%%%%%%%%%%%%%%
\renewcommand{\thesection}{Extended Data Note~\arabic{section}}
\renewcommand{\thesubsection}{\arabic{section}.\arabic{subsection}}
\renewcommand{\tablename}{\hspace{-1mm}}
\renewcommand{\thetable}{Extended Data Table~\arabic{table}}
\renewcommand{\figurename}{\hspace{-1mm}}
\renewcommand{\refname}{\vspace{-8mm}}
\renewcommand{\theequation}{E-\arabic{equation}}
\renewcommand{\thefigure}{Extended Data Figure \arabic{figure}}
\setcounter{figure}{0}    
\setcounter{equation}{0}    
\setcounter{table}{0}

%%%%%%%%%%%%%%%%%%%%%%%%%%%%%%%%%%%%%%%%%%%%%%%%%%%%%%%%%%%
\section*{Methods}\label{section:method}
%%%%%%%%%%%%%%%%%%%%%%%%%%%%%%%%%%%%%%%%%%%%%%%%%%%%%%%%%%%
\subsection*{Experimental Setup}\label{section:setup}
%%%%%%%%%%%%%%%%%%%%%%%%%%%%%%%%%%%%%%%%%%%%%%%%%%%%%%%%%%%

\begin{figure}[h!]
\centering
\includegraphics[width=0.9\textwidth]{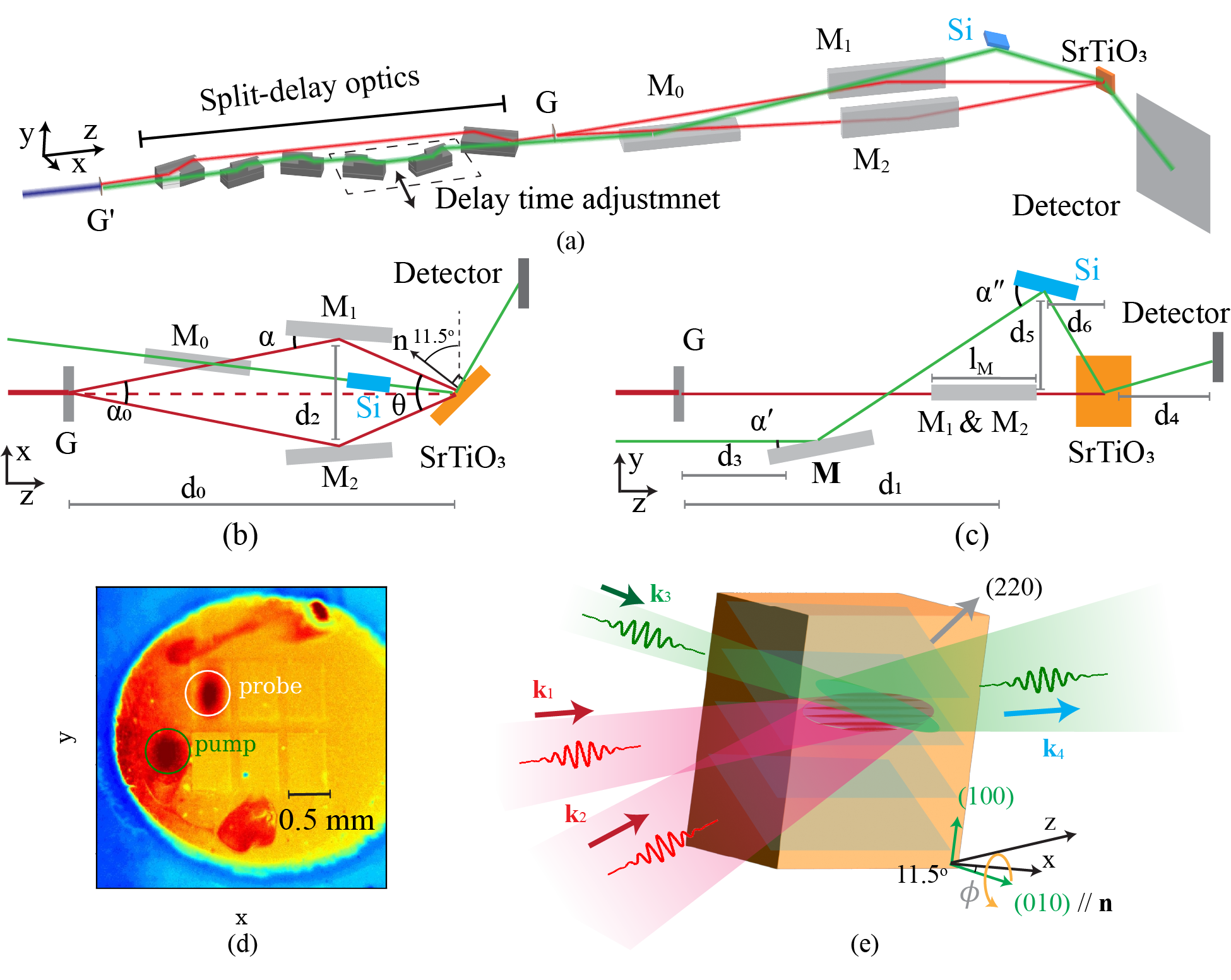}
\caption{
$\vert$~ \textbf{Schematics of experimental setup}:
(a) The layout of X-ray optics, sample and detector inside the experiment hutch.
(b) and (c) are X-ray trajectorys in the $x$-$z$ and $y$-$z$ plane downstream the SDO.
(d) Pump and probe pulse position on grating, $G$. Image taken with cameras during the experiment.
The rectangles are independent gratings on the same diamond substrate. 
(e) Relative orientation of X-ray pulses and the sample.
The transparent parallelograms inside the orange cubic indicates the atomic plane for the (220) reciprocal lattice of STO. 
}\label{fig:whole setup}
\end{figure}

The layout of the experimental setup inside the XPP hutch at LCLS is shown in \ref{fig:whole setup} (a).
The sizes and distances of different components are scaled for better visualization.
Downstream the XPP monochromator \cite{zhu2014performance}, the X-ray is focused with a compound refractive lens (CRL) with a focusing length of 15~m.
The sample is located upstream of the focal plane.
After the CRL, the incident X-ray pulse is split by the split-delay optics (SDO) \cite{li2021generation} into pump and probe pulses as represented by the red and green trajectories in \ref{fig:whole setup} (a).
The SDO controls the relative position and delay between the pump and probe pulse at its exit.
On the diamond grating downstream of the SDO, G, the relative position between the pump and probe pulse is shown in \ref{fig:whole setup} (d). 
The probe beam is separated from the pump beam along the y axis and x axis so that it can avoid grating diffraction with G and be reflected by $M_0$.
One can perform a delay time scan between 0 to 13~ps through continuous motion of the crystals of SDO.
The mechanism and performance of the SDO are characterized in another study \cite{li2021generation}.

Downstream the SDO, the 2D visualization of the X-ray trajectory in the x-z and y-z plane is shown in \ref{fig:whole setup}(b) and (c).
The grating $G$ is a square transmission diamond phase grating with a grating period of $1~\mu$m and the diffraction efficiency into $\pm1^\text{st}$ diffraction orders is both $28\%$.
Within the x-z plane as shown in \ref{fig:whole setup} (b), the angular separation between the $\pm1$ diffraction order of $G$ is $\alpha_0=0.0145\degree$.
The $M_0$, $M_1$ and $M_2$ are Rh-coated total reflection mirrors, each with a length of $l_M=40~mm$.
The $M_1$ and $M_2$ are $d_1\approx5.9$~m downstream $G$, and the spatial separation between two pump pulses at $M_1$ and $M_2$ is $d_2\approx 1.5~\text{mm}$.
The two pump pulses have the same grazing incident angle $\alpha$ with respect to the $M_1$ and $M_2$.
The symmetry of the X-ray trajectory within the $x$-$z$ plane guarantees the synchronization of the two pump pulses.

\begin{figure}[bht!]
\centering
\includegraphics[width=0.85\textwidth]{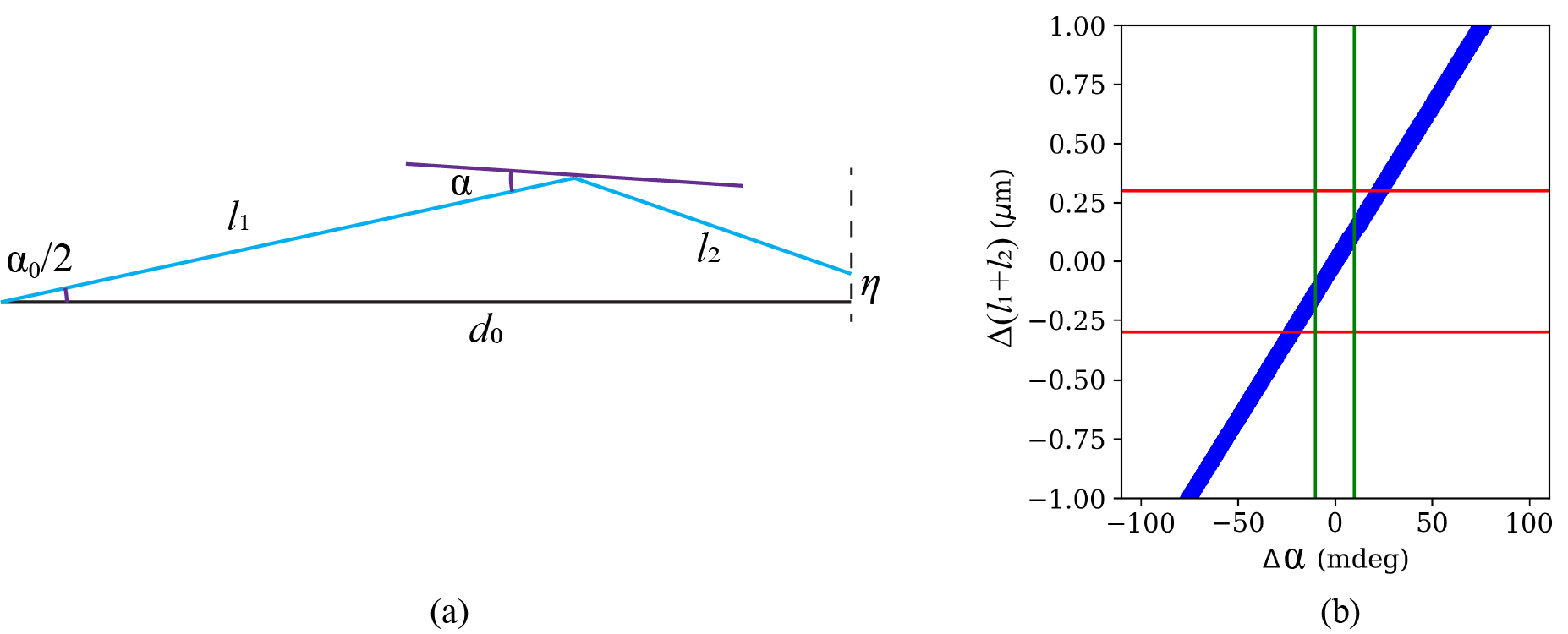}
\caption{$\vert$~\textbf{Trajectory of pump pulse}:
(a) shows the trajectory of one pump pulse between $G$ and the sample. 
For a given grating-sample distance $d_0=6$~m, grating diffraction angle $\alpha_0 / 2$, beam position on the sample plane $\eta$, and mirror angle $\alpha$, one can uniquely determine the total path length $l_1+l_2$. 
In the ideal case, $\eta=0$, and $\alpha$ is determined by the targeted $\Lambda_\text{TG}$.
(b) shows the dependence of $l_1 + l_2$ of $\alpha$ around the ideal value of $0.15\degree$ for $\eta\in\left[-2,~2\right]~\mu$m.
The green vertical lines mark the boundary of $\Delta\alpha = \pm 0.01\degree$.
The red horizontal lines mark the boundary of $\Delta(l_1+l_2) = \pm 0.3~\mu$m.
Therefore, a beam overlap accuracy of $|\Delta\eta| \leq 2~\mu$m, and a mirror angle uncertainty of $|\Delta \alpha| \leq 0.01\degree$ guarantees a synchronization better than 1~fs between the two pump pulses. 
}\label{fig:ray_tracing}
\end{figure}

\begin{figure}[bht!]
\centering
\includegraphics[width=0.9\textwidth]{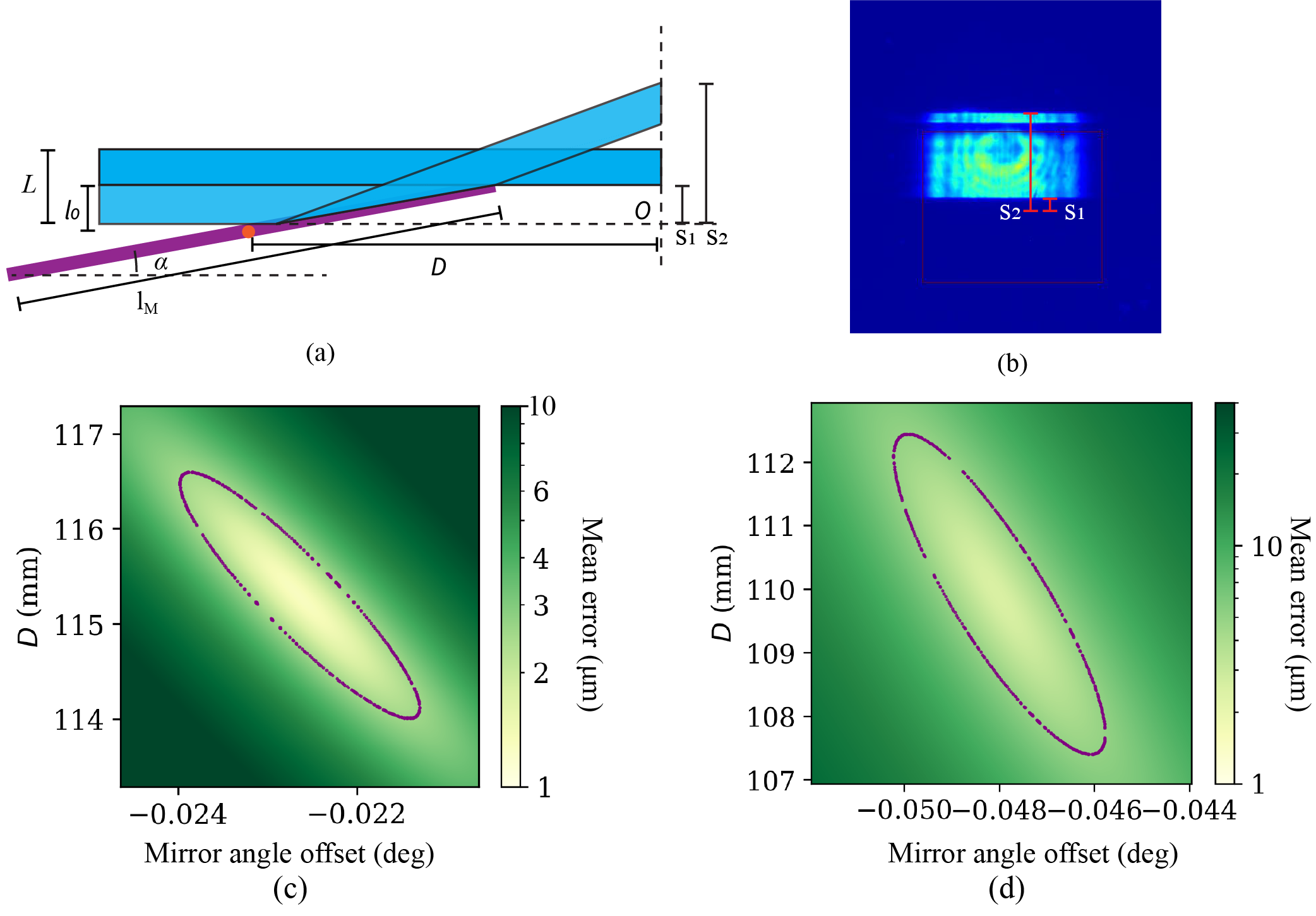}
\caption{
$\vert$~\textbf{Mirror angle calibration}:
(a) shows the schematics of X-ray partial reflection with the mirror. 
The rotation center, marked with the red dot, lies at the center of the mirror reflection surface. 
The lower edge of the x-ray pulse passes through the mirror rotation center and intersect with the sample plane at $O$.
The X-ray beam size is $L$, and the disitance between beam center and rotation center is $l_0\approx L/2$.
For a grazing incidence angle $\alpha$, part of the X-ray is reflected by the mirror.
(b) shows the X-ray beam profile on the sample plane.
The upper edge of the reflected X-ray is separated from $O$ by $S_2$, while the lower edge of the un-reflected X-ray is separated from $O$ by $S_1$.
(c) and (d) show the mean error between simulated and measured $S_2-S_1$ for $M_1$ and $M_2$ respectively over a range of mirror-sample distance $D$, and motor angle offset $\alpha_\text{off}$.
The measurement data are documented in  \ref{table: mirror data}.
The purple ellipse in (c) and (d) marks the region with a value smaller than twice the global minimum. 
The projection of the ellipse gives the uncertainty of $D$ and $\alpha_\text{off}$.
The mirror angle offset $\alpha_\text{off}$ is the difference between the grazing incident angle $\alpha$ and the recorded motor feedback value $\alpha_f =\alpha +\alpha_\text{off}$.
The accuracy of $\alpha_f$ is $0.000765\degree$ according to the spec.
Therefore the uncertainty of $\alpha$ equals to that of $\alpha_\text{off}$, which determined to be $0.003\degree$.
}\label{fig:mirror_align}
\end{figure}

The ray-tracing modeling is utilized to analyze the influence of the installation and alignment imperfections. 
For a given $d_0=6$~m and position accuracy of $\Delta \eta \leq 2~\mu$m of X-ray on the sample plane as shown in \ref{fig:ray_tracing}(a), the total path length variation is less than $0.3~\mu$m or 1~fs if the uncertainty of $\alpha$ is less than $0.01\degree$ as shown in \ref{fig:ray_tracing}(b). 
Details of the ray-tracing calculation are summarized in Supplementary Note 1.1.
The angle uncertainty of $M_1$ and $M_2$ is determined through partial reflections as shown in \ref{fig:mirror_align}(a).
For a series of mirror angles, we measure edge separations $s_2-s_1$ as summarized in \ref{table: mirror data} for $M_1$ and $M_2$.
The derivation of the uncertainty of $0.003\degree$ for $\alpha$ is summarized in Supplementary Note 1.2 and \ref{fig:mirror_align}(b) and (c), which guarantees the sub-femtosecond synchronization of the two pump pulses.

On the sample, the crossing angle between two pump pulses is $\theta\equiv 4\alpha - \alpha_0$, and $\Lambda_\text{TG} \equiv \lambda_0 \left(2 \sin\left(\theta /2\right)\right)^{-1}$.
For $\Lambda_\text{TG} = 11.8~nm$, $\theta = 0.615\degree \pm 0.006\degree$, while for $\Lambda_\text{TG} = 9.2~nm$, $\theta = 0.785\degree \pm 0.006\degree$.

The probe pulse, shown in \ref{fig:whole setup} (b) with green trajectories, is first reflected within the y-z plane by mirror $M_0$.
The distance between $M_0$ and $G$ is $d_3\approx 2~m$. 
The incident angle is $\alpha^\prime = 0.1\degree \pm 0.01\degree$.
After 4~m of propagation, the probe pulse is higher in y than $M_1$ and $M_2$ and then reflected down to the sample with a silicon (111) Bragg reflection mirror, indicated with Si in \cref{fig:whole setup} (a) to (c).
The vertical distance along the y axis between Si and the sample is $d_5\approx20 mm$ and the separation along the z axis is roughly $d_6\approx 7~mm$.
The Si satisfies the Bragg condition for the probe pulse, i.e., $\alpha^{\prime\prime}=11.64\degree$ for 9.8~keV.
By adjusting the position of Si along the x, y, and z axes, we overlap the probe pulse with the two pump pulses.

As shown in \ref{fig:whole setup} (b) and (e), the normal direction of the sample surface is within the x-z plane and has an angle of $11.5\degree$ with respect to the x axis.
The sample is rotating around its normal direction, corresponding to the reciprocal lattice vector (010) of the STO crystal.
The STO (220) reciprocal lattice vector, $\textbf{H}$, is close to the Bragg condition of the probe pulse.
The atomic planes corresponding to the STO (220) reciprocal lattice vector are shown with semi-transparent gray parallelograms in \ref{fig:whole setup} (e).

We use a Jungfrau 1M detector \cite{mozzanica2018jungfrau} to measure the scattering X-ray intensity from the sample. 
The detector distance with respect to the sample along the z axis is $120~mm$, which is calibrated through the \ce{LaB6} powder diffraction as shown in \ref{fig:qCalibration}.
The reduced wave-vector $\textbf{q}$-map of the detector region of interest is shown in \ref{fig:reducedQMap}.
Details of the $\textbf{Q}$ calibration and the derivation of the $\textbf{q}$-map are summarized in Supplementary Note 1.3.

\subsection*{Data Analysis}\label{section:fitting}

According to \cref{eq:final r}, we derive $v$, $\tau_0$, and $\tau_1$ from the measurement shown in \cref{time_trace_and_spec} by fitting the data with the following functional form:
\begin{equation}
    r\left(\pm \textbf{k}_\text{TG}, \tau\right)=A+F\left(e^{-\frac{\tau}{\tau_0}}-e^{-\frac{\tau}{\tau_1}}\cos{\left(k_\text{TG}vt\right)}\right)^2  \label{eq:eta function}.
\end{equation}
where $A\approx1$ is the relaxation of the requirement of $r(\pm\textbf{k}_\text{TG}, 0)=1$ due to noise in the measurement, 
and $F$ accounts for the ratio between the scattering background without XTG pumps and the modulation of the scattering intensity triggered by pump pulses.
The fitting result is summarized in \ref{table:TG fitting}.

\begin{figure}[h!]
    \centering
    \includegraphics[width=0.65\linewidth]{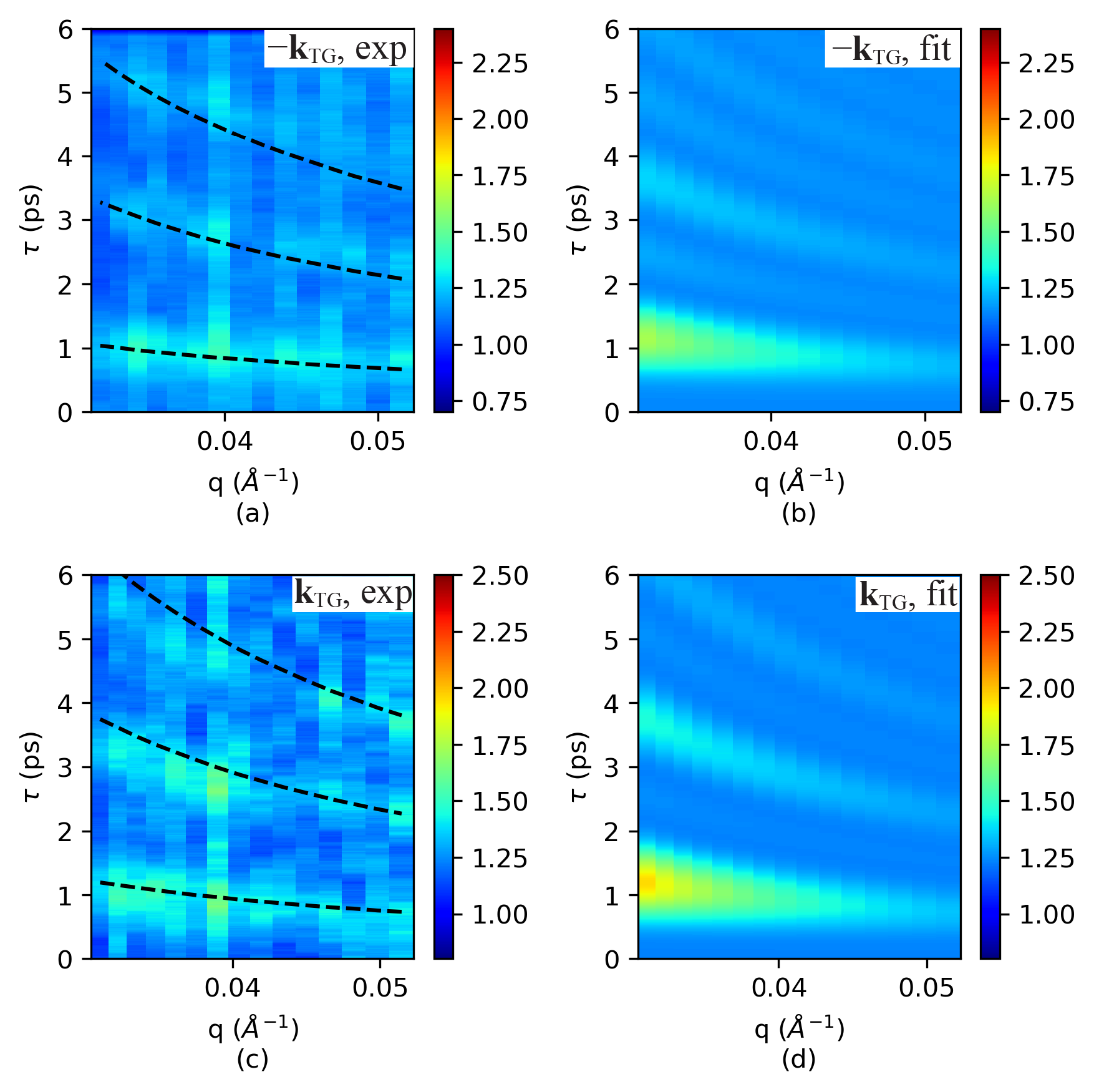}
    \caption{$\vert~$
    \textbf{X-ray diffuse scattering fitting}:
    (a) and (c) are measured $r(\textbf{q},\tau)$ respectively for $-\textbf{k}_\text{TG}$ and $\textbf{k}_\text{TG}$, excluding TG peaks.
    (b) and (d) are fitting result with \cref{eq:fit function} for (a) and (c) respectively.
    The black dashed lines in (a) and (c) are guiding lines, indicating the temporal maximum for each $q$ with the fitting result.}
    \label{fig:amplification}
\end{figure}

For $\textbf{q} \neq \pm \textbf{K}_\text{TG}$, the signal-to-noise ratio of $r(\textbf{q},\tau)$ in \cref{det_img_and_time_trace}(c) and (d) is not high enough for reliable determination of $\tau_0$ and $\tau_1$ from \cref{eq:strain fields} and \cref{eq:diffuse scattering}.
Therefore, we only extract the spatial extension $\sigma$ of the lattice modulation associated with each single X-ray photon absorption site, with the following fitting functional form:
\begin{equation}
    r\left(\textbf{q}, \tau\right)=A+Fe^{-\frac{\sigma^2q^2}{2}}\left(e^{-\frac{\tau}{\tau_0}}-e^{-\frac{\tau}{\tau_1}}\cos{\left(qv\tau\right)}\right)^2\left(\mathbf{Q}\cdot\hat{\mathbf{q}}\right)^2 \label{eq:fit function},
\end{equation}
We have dropped the $q$ dependence of $F$ in this analysis due to the limited $q$ range of our measurement and have used $v$, $\tau_0$ and $\tau_1$ from the fitting with $r\left(\pm \textbf{k}_\text{TG}, \tau\right)$ at XTG diffraction peaks.
The comparison between the measurement and fitting result is shown in \ref{fig:amplification} and the fitting parameters are summarized in \ref{table:site size}.
The characteristic size of the excitation site measured in $\sigma=1.4\pm0.1$~nm is compatible with results from previous studies \cite{huang2024Nanometer}.

\section*{Data availability statement} 
The data supporting the findings of the study are included in the main text and supplementary information files.
Source data have been deposited in the Stanford Digital Repository~\url{https://purl.stanford.edu/zy110hn7917}.
Additional data are available from the corresponding author upon request.

\section*{Code availability} 
Code for producing figures in the main text, extended data, and supplementary information files has been deposited in the Stanford Digital Repository~\url{https://purl.stanford.edu/zy110hn7917}.
Additional code to access and process raw experiment data is available from the corresponding author upon request.

\newpage

\section*{References}
\bibliography{sn-bibliography}% common bib file

% Copy-pasted from my old paper. Need update
\section*{Acknowledgement}
% Haoyuan, Matthias
%Financial support from the U.S. Department of Energy, Office of Science under DOE (BES) Awards DE-SC0022222 (H.L., M.I.) is gratefully acknowledged. 
%HL was also supported by the U.S. Department of Energy (DOE), Office of Science, Office of Basic Energy Sciences, Division of Chemical Sciences, Geosciences, and Biosciences, Gas-Phase Chemical Physics Program award number FWP-100778 granted to the SLAC Chemical Science Division.
Financial support from the U.S. Department of Energy, Office of Basic Energy Sciences, Gas-Phase Chemical Physics Program with Awards DE-SC0022222 and FWP-100778 (H.L., M.I.) is gratefully acknowledged.
%AM, NB, PRM, ZZ, KAM
The contribution by N.B., P.R.M., Z.Z., K.A.N., and A.A.M. was supported by  the Department of Energy, Office of Science, Office of Basic Energy Sciences, under award number DE-SC0019126.
% LZ, MT, DAR.
L.Z., M.T., and D.A.R. were supported by the U.S. Department of Energy, Office of Science, Office of Basic Energy Sciences through the Division of Materials Sciences and Engineering under Contract No. DE-AC02-76SF00515.
% C.O.
The contribution by C.O. was supported by the AMOS program within the Chemical Sciences, Geosciences, and Biosciences Division, DOE, BES, DOE.
% Y.S., S.Song, T.S., N.W., S.She, M.N., M.D.B., V.E., J.B.H. and D.Z.
The contribution by Y.S., S.Song, T.S., N.W., S.She, M.N., M.D.B., V.E., J.B.H. and D.Z. was supported by the U.S. Department of Energy, Office of Science, Office of Basic Energy Sciences under Contract No. DE-AC02-76SF00515. 
% LCLS
Use of the Linac Coherent Light Source (LCLS), SLAC National Accelerator Laboratory, is supported by the U.S. Department of Energy, Office of Science, Office of Basic Energy Sciences under Contract No. DE-AC02-76SF00515.

%%%%%%%%%%%%%%%%%%%%%%%%%%%%%%%%%%%%%%%%%%%%%%%%%%%%%%%%%%%

\section*{Author Contribution}

H.L. : Data curation, Formal analysis, Investigation, Methodology, Writing -- original draft, Writing -- review \& editing; 
N.W.: Data curation, Formal analysis, Investigation;
L.Z. : Data curation, Formal analysis, Investigation;
S.Song : Project administration, Data curation, Investigation, Methodology;
Y.S. : Data curation, Investigation, Methodology;
M.N. : Investigation, Resources; 
T.S. : Data curation, Investigation, Methodology;
D.H. : Data curation, Formal analysis, Investigation;
S.D. : Data curation, Formal analysis, Investigation;
M.B. : Investigation, Methodology;
V.E. : Investigation, Methodology, Software;
S.She : Data curation, Investigation, Methodology;
C.C. : Data curation, Investigation, Methodology;
J.V. : Investigation, Resources; 
C.D. : Investigation, Resources; 
N.B. : Data curation, Formal analysis, Investigation, Resources;
P.M. : Data curation, Formal analysis, Investigation;
Z.Z. : Investigation, Resources;
M.I. : Funding acquisition, Supervision, Writing - Review \& Editing.
M.T. : Formal analysis, Investigation, Methodology;
K.N. : Supervision, Investigation, Resources; 
J.H. : Supervision, Investigation; 
A.M. : Formal analysis, Investigation, Methodology, Writing -- original draft, Writing -- review \& editing; 
L.F. : Formal analysis, Investigation, Methodology, Writing -- original draft, Writing -- review \& editing; 
S.T. : Formal analysis, Investigation, Methodology, Writing -- original draft, Writing -- review \& editing; 
D.R. : Formal analysis, Investigation, Methodology, Writing -- original draft, Writing -- review \& editing; 
D.Z. : Formal analysis, Investigation, Methodology, Resources, Funding acquisition, Project administration, Supervision, Writing -- original draft, Writing - Review \& Editing.

\section*{Competing Interest} 
The authors declare no competing interests.

\newpage

\newpage

\newpage

\newpage

\newpage

\textbf{\begin{figure}[bht!]
    \centering
    \includegraphics[width=\linewidth]{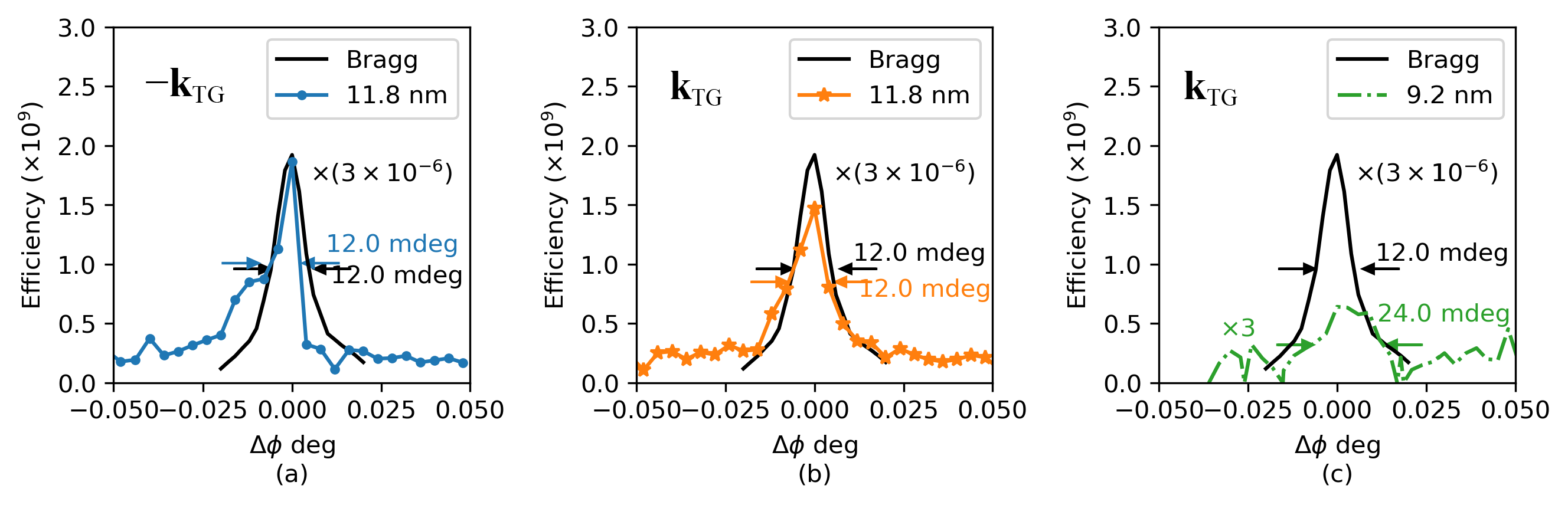}
    \caption{$\vert~$
    \textbf{Angular dependence of efficiency}:
    (a), (b) and (c) are angular dependence of the diffraction efficiency measured for the Bragg peak (black) and the TG peaks.
    For detailed comparison, the horizontal axis $\Delta\phi$ is the relative angle with respect to the peak position of each curve. 
    The number in the legend shows the corresponding XTG wavelength $\Lambda_\text{TG}$.
    The $\pm\textbf{k}_\text{TG}$ on the top left corner indicates the angular relation of the TG peak with respect to the Bragg peak and is the same as that in \cref{det_img_and_time_trace}.
    }
    \label{fig:area}
\end{figure}
}

\newpage

\begin{figure}[bht!]
\centering
\includegraphics[width=0.8\textwidth]{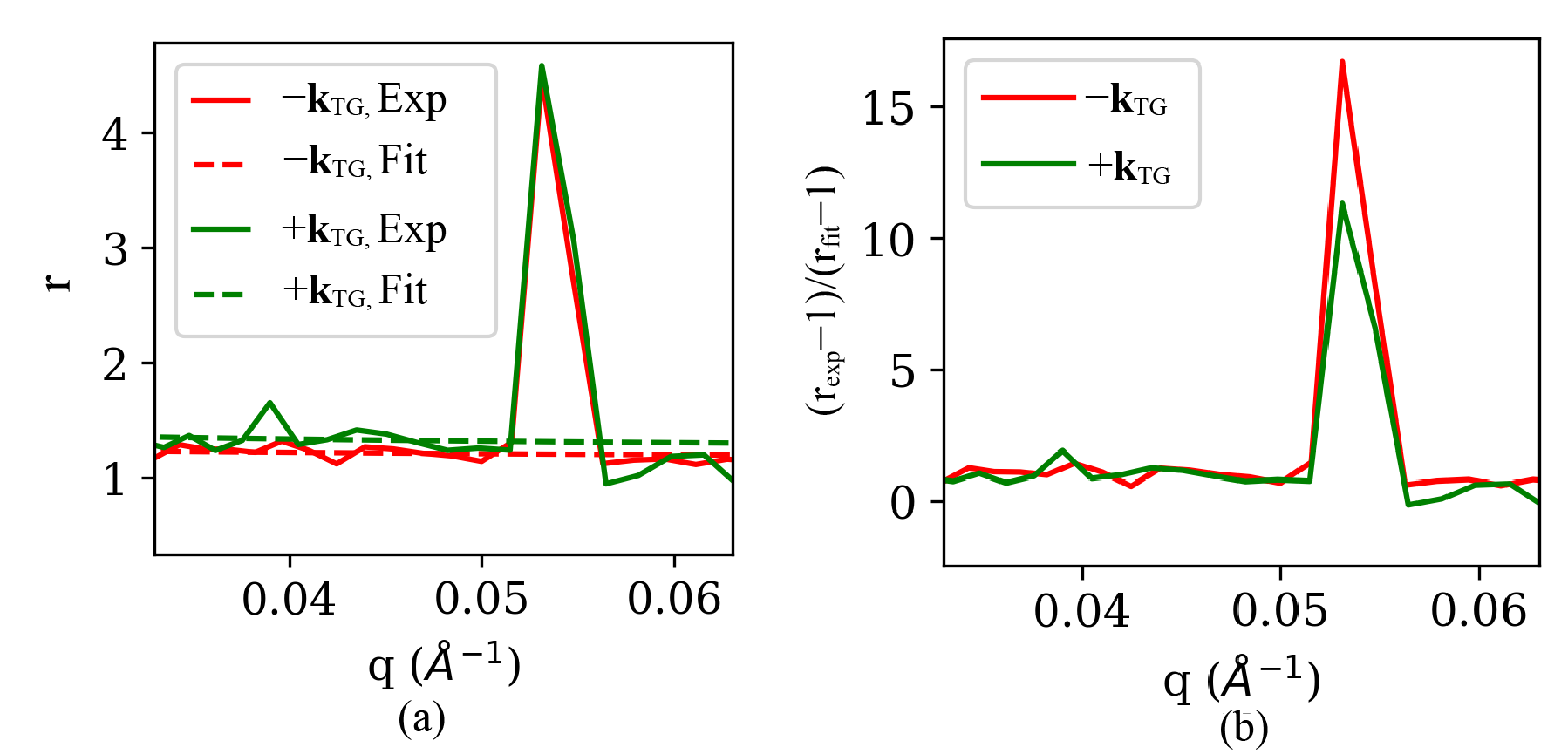}
\caption{\textbf{XTG-enhanced intensity modulation}:
In (a), the solid lines are line cut of (a) along lines of first maxima in $\tau$ for each $q$.
The dashed lines are linear fit of the corresponding solid lines for $q<0.05\text{\AA}^{-1}$.
(b) is the ratio between the measured intensity modulation maxima $r_\text{exp}-1$ shown in solid lines in (a), versus the linear fit shown in dashed lines $r_\text{fit}-1$ in (a).
}\label{fig:enhancement factor}
\end{figure}

\newpage

\begin{figure}[bht!]
    \centering
    \includegraphics[width=0.35\linewidth]{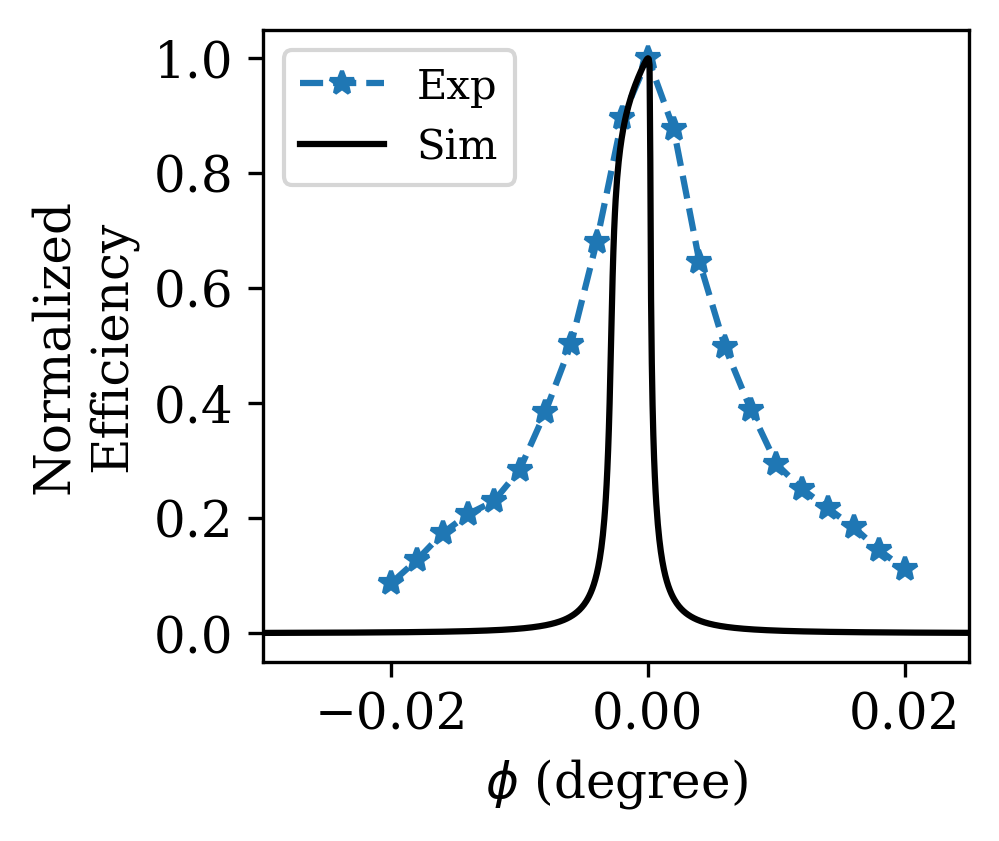}
    \caption{$\vert~$
    \textbf{Comparison of Bragg diffraction efficiency of simulation versus measurement}: 
    The blue star are measured STO (220) Bragg diffraction efficiency for different sample rotation angle $\phi$.
    The black curves are simulation results assuming a perfect crystal within the X-ray illuminated region.
    }
    \label{fig:rocking_simulation}
\end{figure}

\newpage

\begin{figure}[bht!]
\centering
\includegraphics[width=0.8\textwidth]{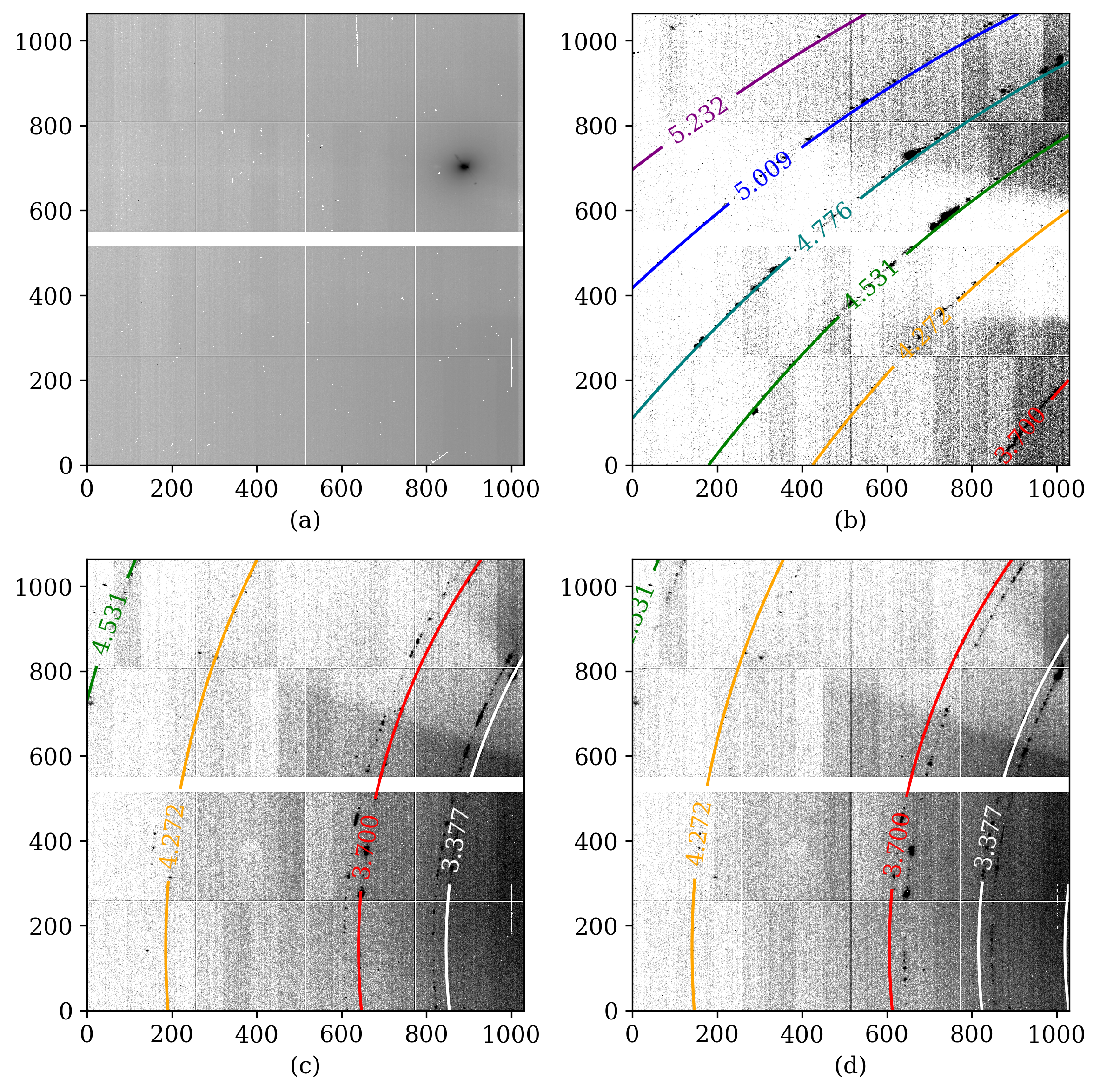}
\caption{$\vert$~
\textbf{Detector $Q$ map calibration}:
(a) Example of the scattering intensity of the Bragg peak from \ce{SrTiO3} measured with the area detector in our experiment.
The axis labels are the pixel indexes.
(b), (c) and (d) show the \ce{LaB6} power scattering ring from the probe and both pump pulses, overlapped with the documented ring position with the calibrated detector position and distance.
The numbers indicate the angular wavenumber measured in $\text{\AA}^{-1}$ for the documented ring position for \ce{LaB6}.
}\label{fig:qCalibration}
\end{figure}

\newpage

\begin{figure}[bht!]
\centering
\includegraphics[width=0.9\textwidth]{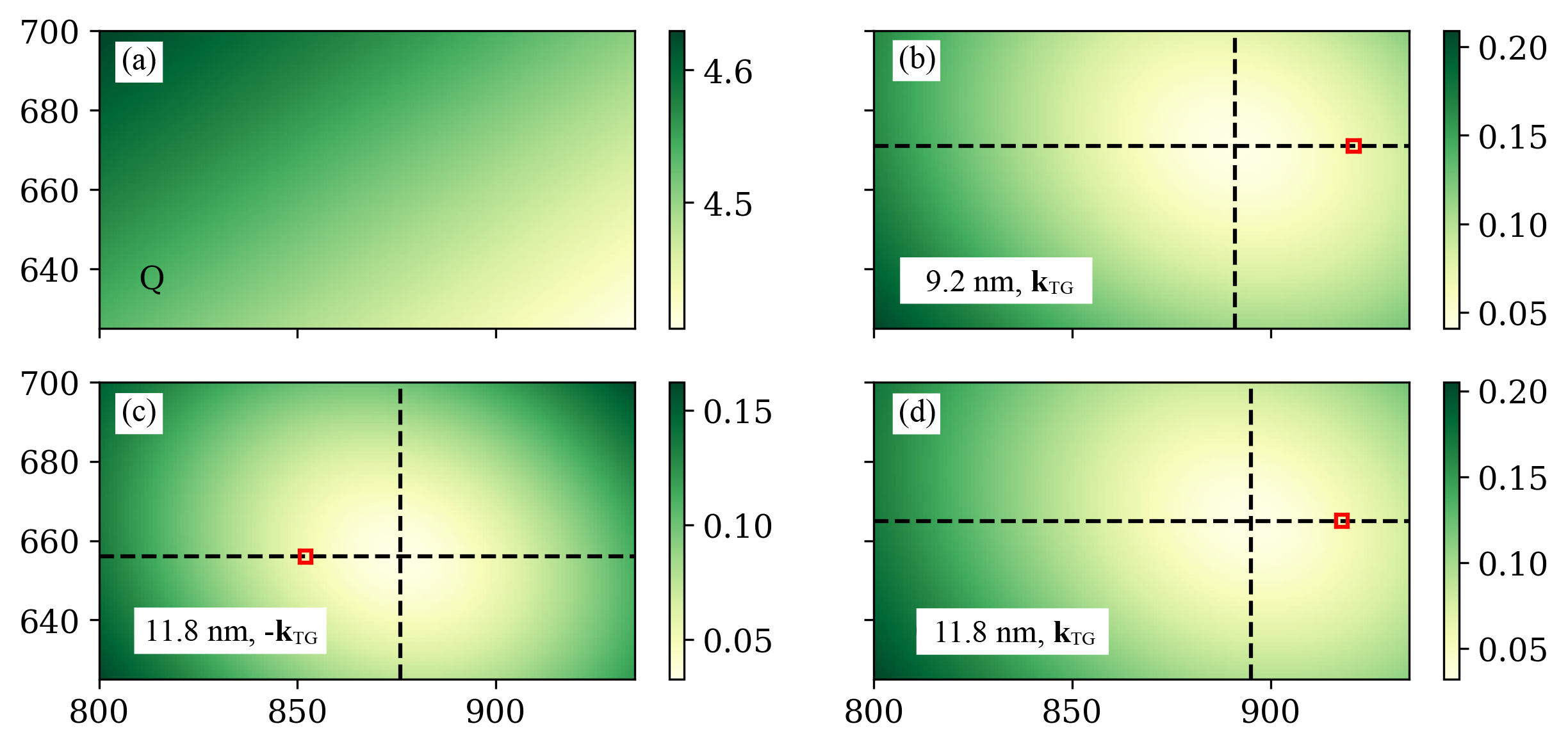}
\caption{$\vert$~
\textbf{Detector $q$ map of the region of interest}:
(a) The $Q$ value for each pixel within the region of interests containing the Bragg peak and TG peaks. The axis labels are pixel indexes as shown in \ref{fig:qCalibration}. The color map has unit of \AA$^{-1}$.
(b), (c) and (d) are $q$ maps of TG measurements for 
$\Lambda_\text{TG}=9.2$ and $11.8$~nm.
The black dashed lines are guiding lines for for minimal $q$ in each plot. 
The red box is the location of the TG peak, and the color map is $q$ measured in \AA$^{-1}$.
}\label{fig:reducedQMap}
\end{figure}

\newpage

\begin{table}[bht!]
\centering
\begin{tabular}{|c|c|c|c|c|c|c|c|} 
\hhline{|--------|}
\multirow{2}{*}{$M_1$} & $S_1-S_1$~($\mu$m) & 81.2227 & 274.2402 & 355.9842 & 464.0974 & 540.5686 & 659.23 \\ 
\hhline{|~|-------|}
 & Motor angle & $0.05\degree$ & $0.1\degree$ & $0.12\degree$ & $0.15\degree$ & $0.17\degree$ & $0.2\degree$ \\ 
\hhline{|--------|}
\multirow{2}{*}{$M_2$} & $S_1-S_1$~($\mu$m) & 184.587 & 384.995 & 588.029 & 788.437 & 469.365 & 667.134 \\ 
\hhline{|~|-------|}
 & Motor angle & $0.05\degree$ & $0.1\degree$ & $0.15\degree$ & $0.2\degree$ & $0.12\degree$ & $0.17\degree$ \\
\hhline{|--------|}
\end{tabular}
\caption{$\vert~$\textbf{Mirror angle calibration data}: The seperation $S_2-S_1$ between edges of X-ray beam profile for different motor angles for $M_1$ and $M_2$ as shown in \ref{fig:mirror_align}(a) and (b). The motor angle in this table is the feedback value of the motor controller. \label{table: mirror data}}
\end{table}

\newpage

\begin{table}[bht!]
\begin{tabular}{|l|l|l|l|l|}
\hline
    & A  (a.u.)             & $\mathcal{F}$   (a.u.)  & $\sigma^2$ (nm$^2$)  \\ \hline
$-\textbf{k}_\text{TG}$ & $1.15\pm0.07$  & $0.020\pm0.081$    & $2.04\pm0.02$   \\ \hline
$\textbf{k}_\text{TG}$ &  $1.21\pm0.02$  & $0.030\pm0.008$    & $2.00\pm0.01$   \\ \hline
\end{tabular}
\caption{$\vert~$\textbf{Fitting result for diffuse scattering}: The fitting parameter of \cref{eq:fit function} with in the $q$-$\tau$ region specified in \ref{fig:area}.
}\label{table:site size}
\end{table}

\newpage
\begin{table}[bht!]
\begin{tabular}{|l|l|l|l|l|l|}
\hline
    & $A$     (a.u.)         & $F$ (a.u.)    & $\tau_0$ (ps) & $\tau_1$ (ps)   & $v$ (km/s)   \\ \hline
$-\textbf{k}_\text{TG}$ & $0.88\pm0.05$  & $0.58\pm0.05$    & $1.8\pm0.4$  & $7.7\pm1.4$ & $8.0\pm0.4$ \\ \hline
$\textbf{k}_\text{TG}$ & $1.01\pm0.08$  & $0.55\pm0.03$    & $3.0\pm1.0$  & $11.9\pm3.6$ & $8.0\pm0.7$  \\ \hline
\end{tabular}
\caption{$\vert~$\textbf{Fitting result for TG dynamics}: The fitting parameter of \cref{eq:eta function} for result shown in \cref{time_trace_and_spec}.
}\label{table:TG fitting}
\end{table}

\end{document}